\documentstyle[preprint,aps,epsf,floats]{revtex}
\begin{document}
\tighten
\def\si{{}^1\kern-.14em S_0}
\def\siii{{}^3\kern-.14em S_1}
\def\diii{{}^3\kern-.14em D_1}
\newcommand{\gsim}{\raisebox{-0.7ex}{$\stackrel{\textstyle >}{\sim}$ }}
\newcommand{\lsim}{\raisebox{-0.7ex}{$\stackrel{\textstyle <}{\sim}$ }}
\def\pislash{ {\pi\hskip-0.6em /} }
\def\pislashsmall{ {\pi\hskip-0.375em /} }
\def\pslash{p\hskip-0.45em /}
\def\nopi{ {\rm EFT}(\pislash) }
\def\Ltwo{ {^\pislashsmall \hskip -0.2em L_2 }}
\def\Lone{ {^\pislashsmall \hskip -0.2em L_1 }}
\def\CQuad{ {^\pislashsmall \hskip -0.2em C_{\cal Q} }}
\def\Czeromone{ {^\pislashsmall \hskip -0.2em C_{0,-1}^{(\siii)} }}
\def\Czerozero{ {^\pislashsmall \hskip -0.2em C_{0,0}^{(\siii)} }}
\def\Ctwomtwo{ {^\pislashsmall \hskip -0.2em C_{2,-2}^{(\siii)} }}
\def\CSDzero{ {^\pislashsmall \hskip -0.2em C_0^{(sd)} }}
\def\CSDtwotwotwo{ {^\pislashsmall \hskip -0.2em C_{2,-2}^{(sd)} }}
\def\CSDzeromone{ {^\pislashsmall \hskip -0.2em C_{0,-1}^{(sd)} }}
\def\CSDzerozero{ {^\pislashsmall \hskip -0.2em C_{0,0}^{(sd)} }}
\def\CSDtwoone{ {^\pislashsmall \hskip -0.2em \tilde C_2^{(sd)} }}
\def\CSDtwotwo{ {^\pislashsmall \hskip -0.2em C_2^{(sd)} }}
\def\CSDzerotwo{ {^\pislashsmall \hskip -0.2em C_{0,0}^{(sd)} }}
\def\LX{ {^\pislashsmall \hskip -0.2em L_X }}
\def\CSDfour{ {^\pislashsmall \hskip -0.2em C_4^{(sd)} }}
\def\CSDfourt{ {^\pislashsmall \hskip -0.2em \tilde C_4^{(sd)} }}
\def\CSDfourtt{ {^\pislashsmall \hskip -0.2em {\tilde{\tilde C}}_4^{(sd)} }}
\def\etasd{\eta_{sd} }

\def\Journal#1#2#3#4{{#1} {\bf #2}, #3 (#4)}

\def\NCA{\em Nuovo Cimento}
\def\NIM{\em Nucl. Instrum. Methods}
\def\NIMA{{\em Nucl. Instrum. Methods} A}
\def\NPB{{\em Nucl. Phys.} B}
\def\NPA{{\em Nucl. Phys.} A}
\def\NP{{\em Nucl. Phys.} }
\def\PLB{{\em Phys. Lett.} B}
\def\PRL{\em Phys. Rev. Lett.}
\def\PRD{{\em Phys. Rev.} D}
\def\PRC{{\em Phys. Rev.} C}
\def\PRA{{\em Phys. Rev.} A}
\def\PR{{\em Phys. Rev.} }
\def\ZPC{{\em Z. Phys.} C}
\def\SJP{{\em Sov. Phys. JETP}}
\def\SJNP{{\em Sov. Phys. Nucl. Phys.}}

\def\FBS{{\em Few Body Systems Suppl.}}
\def\IJMP{{\em Int. J. Mod. Phys.} A}
\def\UJP{{\em Ukr. J. of Phys.}}
\def\CJP{{\em Can. J. Phys.}}
\def\SCI{{\em Science} }
\def\AST{{\em Astrophys. Jour.} }

\preprint{\vbox{
\hbox{ NT@UW-99-20}
}}
\bigskip
\bigskip

\title{Isoscalar M1 and E2  
Amplitudes in $np\rightarrow d\gamma$\ \ \footnote{
The preprint formally known as 
{\it Suppressed Amplitudes in $np\rightarrow d\gamma$}} 
}
\author{Jiunn-Wei Chen$^a$, Gautam Rupak$^a$ and Martin J. Savage$^{a,b}$}
\address{$^a$ Department of Physics, University of Washington, \\
Seattle, WA 98915. }
\address{$^b$ Jefferson Lab., 12000 Jefferson Avenue, Newport News, \\
Virginia 23606.}
\maketitle

\begin{abstract}
The low energy radiative capture process $np\rightarrow d\gamma$
provides a sensitive probe of the two-nucleon system.
The cross section for this process is dominated by the isovector $M1$
amplitude for capture from the $\si$ channel via the isovector magnetic moment
of the nucleon.
In this work we use 
effective field theory to compute the isoscalar $M1$ and isoscalar
$E2$ amplitudes that are strongly suppressed for cold
neutron capture.
The actual value of the isoscalar $E2$ amplitude is expected to be
within $\sim 15\%$ of the value computed in this work.
In contrast, due to the vanishing contribution of the one-body
operator at leading order and next-to-leading order,
the isoscalar $M1$ amplitude is estimated to have a large uncertainty.
We discuss in detail the deuteron quadrupole form factor 
and $\siii-\diii$ mixing.
\end{abstract}

\vskip 2in

\leftline{April 1999; Revised August 1999.}
%
%
%
%
\vfill\eject


The cross section for radiative capture $np\rightarrow d\gamma $ of thermal
neutrons has an important place in nuclear physics as it provides a clear
demonstration of strong interaction physics that is not constrained by
nucleon-nucleon scattering phase shift data alone.
Effective range theory\cite{ERtheory,Noyes}
uniquely
describes the scattering of low-energy nucleons, yet fails to reproduce the
measured cross section of $\sigma ^{{\rm expt}}=334.2\pm 0.5\ {\rm mb}$
(measured at an incident neutron speed of
$\left| {\bf v}\right| =2200$ m/s)\cite{CWCa}
for  $np\rightarrow d\gamma $
at the $10\%$ level.
In the effective field theory appropriate for very low momentum
interactions\cite{CRSa}
(i.e. without pions), this
discrepancy is understood to arise from the omission of a
four-nucleon-one-magnetic-photon operator that enters at the same order as
effective range contributions.
Conventionally, this
discrepancy is attributed to pion-exchange-currents\cite{BrRia,PMRa}.
The cross section for $np\rightarrow d\gamma $ at very low energies is
dominated by the capture of nucleons in the $\si$ state, via
the nucleon isovector magnetic moment,
the amplitude for which we denote by $M1_{V}$.
This particular
amplitude is much larger than other amplitudes for several reasons.
First, initial state interactions give a contribution proportional
to the large scattering length in the $\si$ channel,
$a^{(\si)}=-23.714\pm 0.013~{\rm fm}$.
Second, the  $M1_{V}$ amplitude is proportional to the
nucleon isovector magnetic moment, $\kappa _{1}$,
which is much larger than the nucleon isoscalar magnetic
moment, $\kappa _{0}$, which dictates the size of the one-body
contribution to the isoscalar magnetic amplitude, $M1_{S}$.
Third, the capture from the $\siii$ channel that does 
proceed via the nucleon isoscalar magnetic interaction
(the one-body contribution)
must vanish at zero-momentum transfer as it is the matrix element of the
spin operator between orthogonal eigenstates.
Finally, the electric
amplitudes, $E1_{V}$ for capture from the P-wave
and $E2_{S}$ for capture from the $\siii$
channel, are suppressed by additional powers of nucleon momentum or
photon energy compared to  $M1_{V}$.

While the  $M1_{S}$, $E2_{S}$ and $E1_{V}$ amplitudes are much
smaller than $M1_{V}$, measurements of spin-dependent observables
can determine specific combinations of these amplitudes.
Two such observables are the circular polarization of photons emitted
in the capture of polarized neutrons by unpolarized protons,
and the angular distribution of photons emitted in the capture
of polarized neutrons by polarized protons.
The circular polarization of photons emitted in the forward direction
in the capture of polarized neutrons on unpolarized protons
has been measured to be\cite{Bazh}
$P_{\gamma}^{\rm expt} = -(1.5\pm 0.3)\times 10^{-3}$.
This value is consistent with previous theoretical estimates\cite{BKa}.
An experiment that will measure the angular distribution of
photons emitted in the capture of polarized neutrons on polarized protons
is to be carried out at the ILL reactor facility\cite{nppolexpt}
and results should be available in the near future.

In this work, we calculate the $M1_{S}$ and  $E2_{S}$ isoscalar
amplitudes that contribute to  $np\rightarrow d\gamma$
using the effective field theory (EFT) of nucleon-nucleon interactions
without pions, $\nopi$, as detailed in \cite{CRSa}, using KSW 
power counting\cite{KSW,KSW2}.
A significant amount of progress has been made in the application of EFT to the
two- and three-nucleon systems\cite{PMRa}\cite{Weinberg1}-\cite{threebod} 
during the past few years.
A test of this formalism will be a comparison between these
predictions for the strongly suppressed amplitudes in $np\rightarrow d\gamma$
and the measured experimental asymmetries which constrain them.
Calculations of these suppressed amplitudes using 
an alternative power counting are being performed by Park, Kubodera, 
Min and Rho\cite{PMRsupp}.
Our work  results from a challenge issued by M. Rho for the community
to make predictions for these
amplitudes\cite{PMRsupp}.


The amplitude for low-energy $np\rightarrow d\gamma $ is
\begin{eqnarray}
  T & = &
 i e \ X_{M1_{V}}\ \varepsilon ^{abc}\epsilon _{(d)}^{\ast a}\
{\bf k}^{b}\
\epsilon _{(\gamma )}^{\ast c}\ 
U^{T}_{\rm n} \tau _{2}\tau _{3}{\bf \sigma }_{2}
\ U_{\rm p} 
\ +\ 
  e\  X_{E1_{V}}\  U^{T}_{\rm n}\ \tau_2\tau_3\ {\bf \sigma }_2\ 
  {\bf\sigma}\cdot\epsilon _{(d)}^*
\ U_{\rm p}\ {\bf p}\cdot\epsilon_{(\gamma)}^*
\nonumber \\
& + & e\ X_{M1_{S}} {1\over\sqrt{2}} 
\ U^{T}_{\rm n} \tau _{2}{\bf \sigma }_{2}\left[ {\bf \sigma }\cdot 
{\bf k}\ \epsilon _{(d)}^{\ast }\cdot \epsilon _{(\gamma )}^{\ast }-\epsilon
_{(d)}^{\ast }\cdot {\bf k}\ \epsilon _{(\gamma )}^{\ast }\cdot {\bf \sigma }
\right]\  U_{\rm p}
\nonumber \\
& + & e\ X_{E2_{S}} {1\over\sqrt{2}}
\ U^{T}_{\rm n}\tau _{2}{\bf \sigma }_{2}\left[ {\bf \sigma }\cdot 
{\bf k}\ \epsilon _{(d)}^{\ast }\cdot \epsilon _{(\gamma )}^{\ast }+\epsilon
_{(d)}^{\ast }\cdot {\bf k}\ \epsilon _{(\gamma )}^{\ast }\cdot {\bf \sigma }
-{\frac{2}{n-1}}\sigma \cdot \epsilon _{(d)}^{\ast }\ {\bf k}\cdot \epsilon
_{(\gamma )}^{\ast }\right]\  U_{\rm p}
\ \ \ ,
\label{eq:npamp}
\end{eqnarray}
where we have shown only the lowest partial waves, corresponding to
electric dipole capture of nucleons in a P-wave with amplitude $ X_{E1_{V}}$,
isovector magnetic capture of nucleons in the $\si$ channel with amplitude $
X_{M1_{V}}$,
isoscalar magnetic capture of nucleons in the $\siii$ channel with amplitude $
X_{M1_{S}}$,
and isoscalar electric quadrupole
capture of nucleons in the $\siii$ channel with amplitude $X_{E2_{S}}$.
As we dimensionally
regulate the divergences that appear in the effective field theory 
we keep explicit space-time dependence in the
amplitudes shown in eq.~(\ref{eq:npamp}), with $n$ the number of
space-time dimensions.
$U_{\rm n}$ is the neutron two-component spinor and $U_{\rm p}$ 
is the proton two-component
spinor.
${\bf p}$ is half the neutron momentum in the proton rest frame,
while ${\bf k}$ is the photon momentum.
The  photon polarization vector is 
$\epsilon _{(\gamma )}$, and $\epsilon _{(d)}$ is the deuteron polarization
vector. For convenience, we define dimensionless variables $\tilde{X}$, by 
\begin{eqnarray}
\frac{|{\bf p}| M_{N}}{\gamma ^{2}}X_{E1_{V}}\  &=&i\frac{2}{M_{N}}
\sqrt{\frac{\pi }
  {\gamma ^{3}}}\
\tilde{X}_{E1_{V}}\quad ,\quad
X_{M1_{V}}\ =i\frac{2}{M_{N}}
\sqrt{\frac{\pi }{\gamma ^{3}}}\ \tilde{X}_{M1_{V}}\quad ,\;
\nonumber \\
X_{M1_{S}}\  &=&i\frac{2}{M_{N}}\sqrt{\frac{\pi }{\gamma ^{3}}}\
\tilde{X}_{M1_{S}}\quad ,\quad X_{E2_{S}}\
=i\frac{2}{M_{N}}\sqrt{\frac{\pi }{\gamma^{3}}}\ \tilde{X}_{E2_{S}}
\quad ,
\label{eq:tildedef}
\end{eqnarray}
where $\gamma=\sqrt{M_{N}B}\sim 45.6~{\rm MeV}$
is the deuteron binding momentum, with $B$
the deuteron binding energy.

By measuring certain observables of the $np\rightarrow d\gamma$ process
the four amplitudes
$\tilde{X}_{E1_{V}}$, $\tilde{X}_{M1_{V}}$, $\tilde{X}_{M1_{S}}$, and
$\tilde{X}_{E2_{S}}$, 
can be determined or constrained.
The simplest quantity to measure is the total cross section for the capture
of unpolarized cold neutrons with speed $|{\bf v}|$
by unpolarized protons at rest
(the neutron velocity  ${\bf v}$ is related to the momentum 
${\bf p}$ by
${\bf p}={1\over 2} M_N {\bf v}$,
where relativistic corrections have been neglected).
In terms of the amplitudes given in eq.~(\ref{eq:npamp}) and
eq.~(\ref{eq:tildedef})
the unpolarized cross section is 
\begin{equation}
\sigma =\frac{8\pi \alpha \gamma ^{3}}{M_{N}^{5} |{\bf v}|}
\left[ |\tilde{X}_{M1_{V}}|^{2}
  \ +\ |\tilde{X}_{E1_{V}}|^{2}
  \ +\ |\tilde{X}_{M1_{S}}|^{2}
  \ +\ |\tilde{X}_{E2_{S}}|^{2}
\right]
\ \ \ ,
\label{unpol}
\end{equation}
where $\alpha $ is the fine-structure constant.
The cross section for the capture of cold neutrons is dominated by
$\tilde{X}_{M1_{V}}$ by several orders of magnitude
and therefore a measurement of $\sigma$ does not constrain 
the other three amplitudes.

A spin-polarized neutron beam
incident upon a spin-polarized proton target
enables spin-dependent observables to be measured, even without measuring the
polarization of the out-going photon or deuteron.  
If the protons have polarization $\eta_{\rm p}$ 
and the neutrons have polarization
$\eta_{\rm n}$, along the direction of the incident
neutron momentum, the spin-dependent capture cross section is
\begin{eqnarray}
I_{\eta_{\rm n}\eta_{\rm p}} (\theta)\ =\ 
  {\frac{d\sigma _{\eta_{\rm n}\eta_{\rm p}}}{d\cos \theta }} & = &
\frac{4\pi \alpha \gamma ^{3}}{M_{N}^{5} |{\bf v}|}
\left[
  |\tilde{X}_{M1_{V}}|^{2}(1-\eta_{\rm n}\eta_{\rm p})
\ +\
{1\over 2} |\tilde{X}_{E1_{V}}|^{2}\sin^{2}\theta\  (3+\eta_{\rm n}\eta_{\rm p})
 \right.   \nonumber \\
&  &\left.
  + \ {1\over 2} |\tilde{X}_{M1_{S}}|^{2}\ 
  (2+\eta_{\rm n}\eta_{\rm p}\ \sin^{2}\theta)
+\ {1\over 2} |\tilde{X}_{E2_{S}}|^{2}\ 
(2+\eta_{\rm n}\eta_{\rm p}\ \sin^{2}\theta )
\ \right.   \nonumber \\
& & \left.
  +\ {\rm Re}\left( \tilde{X}_{M1_{S}}^{{}}\tilde{X}_{E2_{S}}^{\ast
}\right) \eta_{\rm n}\eta_{\rm p}(1-3\cos ^{2}\theta )
\ \right.   \nonumber \\
& & \left.
\ -\ 2\sqrt{2}\ {\rm Re}[\tilde{X}_{E1_{V}}\tilde{X}_{E2_{S}}^*]
\eta_{\rm n}\eta_{\rm p}\ \cos\theta \sin^{2}\theta
\qquad\right]
\ \ \ ,
\label{eq:spincross}
\end{eqnarray}
where $\theta$ is the angle between the polarization axis and the direction
of the emitted photon.
Spin-averaging the expression given in eq.~(\ref{eq:spincross})
over the initial nucleon spin states,
$(\eta _{\rm n},\eta_{\rm p})=(\pm 1,\pm 1)$
and integrating over all angles
reproduces the spin independent cross section shown in eq.~(\ref{unpol}).
From this one can define the angular asymmetry, 
\begin{eqnarray}
  S_{\eta_{\rm n}\eta_{\rm p}}(\theta) &\  = \ &
  { I_{\eta_{\rm n}\eta_{\rm p}} (\theta) - I_{\eta_{\rm n}\eta_{\rm p}} (0)
    \over
    I_{\eta_{\rm n}\eta_{\rm p}} (\theta) + I_{\eta_{\rm n}\eta_{\rm p}} (0) }
\ = \
\frac{\sin^{2}\theta }{4 (1-\eta_{\rm n}\eta_{\rm p})}
\left[ S^{(1)}
    \ +\
    \eta_{\rm n}\eta_{\rm p} S^{(2)} \right]
\ \ \ ,
\label{eq:asym}
\end{eqnarray}
where
\begin{eqnarray}
S^{(1)} & = & {  3 |\tilde{X}_{E1_{V}}|^{2} \over 
 |\tilde{X}_{M1_{V}}|^2 }
\nonumber\\
S^{(2)} & = & {\left(  |\tilde{X}_{E1_{V}}|^{2}
      \ +\  |\tilde{X}_{M1_{S}}|^{2}
      \ +\  |\tilde{X}_{E2_{S}}|^{2}
      \ +\ 6 {\rm Re} [\tilde{X}_{M1_{S}}\tilde{X}_{E2_{S}}^*] 
      \ -\ 4\sqrt{2} {\rm Re} [\tilde{X}_{E1_{V}}\tilde{X}_{E2_{S}}^*]
      \cos\theta
      \right)
    \over \ |\tilde{X}_{M1_{V}}|^2
    }
\ \ \ \ .
\label{eq:sdef}
\end{eqnarray}
For systems with high polarization, measurement of 
this angular asymmetry constrains the small amplitudes.
In the expressions for $S^{(1)}$ and $S^{(2)}$ that appear in
eq.~(\ref{eq:sdef}) we have neglected the small $M1_S$, $E2_S$ and $E1_V$
amplitudes in the denominators.

If the polarization of the out-going photon can be measured, then other
spin-dependent observables can be considered.
For a polarized neutron incident upon an unpolarized proton target, 
there is a different cross section for production of right-handed
versus left-handed
circularly polarized photons.
Defining the asymmetry $A^\gamma_{\eta_{\rm n}} (\theta)$ to be the 
ratio of the  difference to the sum of these
cross sections, 
\begin{eqnarray}
  A^\gamma_{\eta_{\rm n}} (\theta) & = &
  \eta_{\rm n} \left[\ 
    \left( P_\gamma (M1)\ +\   P_\gamma (E2) \right)\ \cos\theta
    \ +\   P_\gamma (E1)\ \sin^2\theta
    \right]
\ \ \ ,
\label{eq:gamas}
\end{eqnarray}
where
\begin{eqnarray}
   P_\gamma (M1) & = &
    {\sqrt{2} {\rm Re}[ \tilde X_{M1_V} \tilde X_{M1_S}^*]
     \over  |\tilde X_{M1_V}|^2}
\ \ \ ,\ \ \ 
   P_\gamma (E2)\ =\ 
    {\sqrt{2} {\rm Re}[ \tilde X_{M1_V} \tilde X_{E2_S}^*]
     \over  |\tilde X_{M1_V}|^2}\ \ ,
\nonumber\\
   P_\gamma (E1) & = &  
    { {\rm Re}[ \tilde
      X_{M1_V} \tilde X_{E1_V}^*]\over |\tilde X_{M1_V}|^2}
\ \ \ ,
\label{eq:pdef}
\end{eqnarray}
where we have again neglected the small $M1_S$, $E2_S$ and $E1_V$
amplitudes in the denominators.

The four amplitudes  
$\tilde{X}_{E1_{V}}$, $\tilde{X}_{M1_{V}}$, $\tilde{X}_{M1_{S}}$, and
$\tilde{X}_{E2_{S}}$, 
can be computed with $\nopi$.
Power counting the leading order (LO) versus next-to-leading order (NLO) for a
given amplitude is straightforward and follows the well known power counting
rules\cite{CRSa,KSW,KSW2}.
However, power counting amplitudes relative to each other is not so
straightforward.
The reason for this is that there are two different kinematic scales for the
capture of cold or thermal neutrons --
the photon energy and the momentum of the incident neutron.
While the velocity of the incident neutron is always assumed to be small,
its finite value gives rise to an $E1_V$ amplitude,
which for $|{\bf v}|=2200~{\rm m/s}$ is 
comparable to the subleading $M1_S$ and $E2_S$ amplitudes.


It is convenient to express the $\tilde{X}$ amplitudes as a 
series in powers of  $Q$;
$\tilde{X} =  \tilde{X}^{(-1)}+\tilde{X}^{(0)}+\tilde{X}^{(1)}+\cdots $
where $Q\sim \gamma / m_\pi$
is the small expansion parameter in the theory and superscripts denote
the order in $Q$. 
The isovector $M1$ amplitude  $\tilde{X}_{M1_{V}}$ has been computed with EFT
previously\cite{CRSa,SSWst} up to NLO.
The amplitude starts at $Q^{0}$ in the power counting,  
\begin{equation}
\tilde{X}_{M1_{V}}^{(0)}=\kappa _{1}\left( 1-a^{(\si)} \gamma \right) \quad ,
\end{equation}
where $\kappa _{1}=(\kappa _{p}-\kappa _{n})/2$ is the isovector nucleon
magnetic moment in nuclear magnetons, with $\kappa _{p}=2.79285$,
$\kappa_{n}=-1.91304$.
While naively, $\tilde{X}_{M1_{V}}^{(0)}$ is of order $Q^0$, numerically
$\tilde{X}_{M1_{V}}^{(0)}\sim 20$ due to the large numerical values of both
$\kappa _{1}$ and $a^{(\si)}$.

At order $Q^{1}$
there are contributions to  $\tilde{X}_{M1_{V}}$
from insertions of the effective range parameter and
also contributions from a four-nucleon-one-magnetic operator, described by the
Lagrange density
\begin{eqnarray}
{\cal L} & = & e\  \Lone\  \left( N^{T}P_{i}N\right) ^{\dagger }\left( N^{T}
    \overline{P}_{3}N\right) {\bf B}_{i}\ +\ {\rm h.c.}
\ \ \ ,
\label{counterL1}
\end{eqnarray}
where ${\bf B} =\nabla\times {\bf A}$ is the magnetic field operator.
$P_{i}$  and  $\overline{P}_{i}$ are  the $\siii$ and $\si$
spin-isospin projection operators respectively, with 
\begin{equation}
P_{i}=\frac{1}{\sqrt{8}}\sigma _{2}\sigma _{i}\ \tau _{2}\quad ,\quad 
\overline{P}_{i}=\frac{1}{\sqrt{8}}\sigma _{2}\ \tau _{2}\tau _{i}
\quad .
\label{eq:projdef}
\end{equation}
The NLO contribution to the amplitude is found to be\cite{CRSa,SSWst}
\begin{eqnarray}
\tilde{X}_{M1_{V}}^{(1)} &=&\
{1\over 2} \kappa_1\rho_d \gamma \left( 1-a^{(\si)} \gamma \right) 
  \nonumber \\
&& -{\displaystyle{M_{N}a^{({}^{1}\kern-.14emS_{0})}\gamma ^{2} \over 4\pi }}
\ (\mu -\gamma )(\mu -\frac{1}{a^{({}^{1}\kern-.14emS_{0})}})\left[ \Lone -
{\displaystyle{\kappa _{1}\pi  \over M_{N}}}
\left( 
{\displaystyle{r_{0}^{(\si)}{}^{{}}
    \over \left( \mu -\frac{1}{a^{(\si)}} \right) ^{2}}}
+
{\displaystyle{\rho _{d} \over (\mu -\gamma )^{2}}}
\right) \right] \hspace{0.05cm}\quad ,
\end{eqnarray}
where $r_{0}=2.73\pm 0.03$ fm is the effective range in the $\si$ channel
and  $\rho _{d}=1.764$ fm is effective range in the $\siii$ channel. 
$\mu $ is the renormalization scale, and 
the $\mu$-dependence of $\Lone$ yields a renormalization scale independent
amplitude, by construction\cite{CRSa,SSWst}.
For convenience we choose $\mu =m_{\pi }$.
As $\tilde{X}_{M1_{V}}$ is the dominant amplitude for the capture process, 
$\Lone=7.24~{\rm fm^4}$ from the unpolarized cross section
\cite{CRSa} in eq.~(\ref{unpol}).


The cross section for any finite incident nucleon momentum has a contribution
from isovector $E1$ capture.
Recently, a N$^3$LO calculation of this amplitude has been 
performed\cite{CSbb} for non-zero energy capture.
At N$^3$LO there are contributions from the effective range parameter and 
from P-wave initial-state interactions which are found to be small.
Neglecting the P-wave initial-state interactions, the amplitude is found to be,
up to  N$^3$LO
\begin{eqnarray}
\tilde{X}_{E1_{V}} & = & -\frac{|{\bf p}| M_{N}}{\gamma ^{2}}
\left( 1 + {1\over 2}\gamma\rho_d + {3\over 8}\gamma^2\rho_d^2
+{5\over 16}\gamma^3\rho_d^3\right)
\ \ \ .
\label{E1LO}
\end{eqnarray}
Capture from the P-wave introduces the factor of the external nucleon momentum,
$|{\bf p}|$,
forcing the amplitude to vanish at threshold.
The powers of $\gamma \rho _{d}$ that appears in the amplitude are
consistent with the deuteron S-wave normalization factor
$1/\sqrt{1-\gamma\rho_d}$ that arises in
effective range theory.
For moderate incident momenta, where $|{\bf p}|\sim Q$, the LO $E1_V$ amplitude
is of order $Q^{-1}$, and dominates the isovector $M1_V$ amplitude, which
starts at $Q^0$.
However, for smaller incident momentum, the $E1_V$ amplitude becomes less
important. 
If we take $|{\bf p}|\sim Q^2$, the $E1_V$ and $M1_V$ 
amplitudes are of the same
order in the counting, however,
for the neutron incident velocity of $2200~{\rm m/s}$,
numerically $|{\bf p}|\sim Q^4$.


In the zero recoil limit, 
the matrix element of the nucleon magnetic moment operator
between the deuteron and nucleons in the $\siii$ channel, contributing to 
$M1_{S}$, is the matrix element of the spin operator between orthogonal
eigenstates states of the strong interaction and
thus vanishes.
This leads to 
$\tilde{X}_{M1_{S}}^{(0)}=0$
at LO ($Q^0$) and further,
the contribution from the one-body operator at NLO ($Q^1$)
also vanishes.
However, at NLO there is a contribution from a
four-nucleon-one-photon two-body operator
defined by the Lagrange density\cite{CRSa,KSW2}
\begin{eqnarray}
{\cal L} &=& -e\ \Ltwo\ i\epsilon _{ijk}\left(
N^{T}P_{i}N\right) ^{\dagger }\left( N^{T}P_{j}N\right) {\bf B}_{k}+\text{h.c.}
\ \ \ .
\label{counter2}
\end{eqnarray}
At NLO the deuteron magnetic moment is found to be\cite{KSW2}
\begin{eqnarray}
  \mu_M & = & {e\over 2 M_N} \left(\kappa_p\ +\ \kappa_n
  \ +\ \Ltwo\ { 2 M_N \gamma
    (\mu-\gamma)^2\over\pi}\right) 
  \ \ \ .
\label{eq:deutmag}
\end{eqnarray}
Reproducing the experimentally observed value of the deuteron
magnetic moment requires that, at this order\cite{CRSa,KSW2},
$\Ltwo(m_{\pi })=-0.149\text{ fm}^{4}$, 
which is significantly
smaller than the naively estimated size of $\sim 1$ fm$^{4}$.
This two-body interaction  contributes to
$\tilde{X}_{M1_{S}}^{(1)}$, and at NLO we find
\begin{equation}
\tilde{X}_{M1_{S}}^{(1)}=
\sqrt{2} \  {\displaystyle{M_{N}\gamma  \over 2\pi }}
\hspace{0.05cm}\Ltwo (\mu -\gamma )^{2}
\quad .
\label{eq:l2amp}
\end{equation}
While formally the leading contribution,
the smallness of $\Ltwo$ suggests that
the contribution given in eq.~(\ref{eq:l2amp}) might not dominate over
higher order terms,
and the $M1_{S}$
amplitude might not be
predicted well by $\nopi$ at this order.

To make this more concrete, one can imagine a higher dimension
four-nucleon-one-photon local operator that gives rise to a 
contribution of the form
\begin{eqnarray}
  \Delta\tilde X_{M1_S} & \sim & \sqrt{2}
{\displaystyle{M_{N}  \over 2\pi }}
\ \LX\ (\mu-\gamma)^3
\ \left[ (p^2+\gamma^2) +  (p^{\prime 2}+\gamma^2) \right]
\ \ \ ,
\label{eq:highermat}
\end{eqnarray}
between states with nucleon momentum $p$ and $p^\prime$,
since $\LX\sim (\mu-\gamma)^{-3}$.
This object makes a vanishing contribution to the magnetic moment of the
deuteron, while making a non-zero contribution to the rate for 
capture from the
$\siii$ channel
\begin{eqnarray}
 \Delta \tilde{X}_{M1_{S}}& \sim & \sqrt{2}
{\displaystyle{M_{N}\gamma^2  \over 2\pi }}
\ \LX\ (\mu-\gamma)^3\
\ \ \ .
\label{eq:higherdim}
\end{eqnarray}
One naively expects $\LX\sim 1~{\rm fm^6}$, which would make such a
contribution approximately $60 \% $ of the amplitude in
eq.~(\ref{eq:l2amp}).
This relatively large uncertainty in the $M1_S$ matrix element is
consistent with previous calculations of this quantity\cite{Bazh},
and particularly the
most recent (preliminary) work of Park, Kubodera, Min and Rho\cite{PMRsupp},
where they find that different
treatments of the short-range component of the interaction
leads to an approximate $60\% $ uncertainty.
At higher orders, there is a contribution from the one-body operator
due to the finite energy release of the capture process.
Naively, this contribution is much smaller than the expected contribution from
higher dimension operators, as estimated in eq.~(\ref{eq:higherdim}),
and so we do not consider it further.


The $E2_S$ amplitude is dominated by local operators that convert 
$\siii$ states to $\diii$ states and vice versa.
However, the relatively slow convergence in this channel requires that the
calculation be performed to higher orders 
so that  a meaningful  estimate of uncertainties is possible.
In previous works\cite{CRSa} the deuteron quadrupole form factor and 
the $\siii-\diii$ mixing parameter
$\overline{\varepsilon}_1$\cite{stapp,nijmegen}
were computed up to NLO. 
Presently, we compute $\overline{\varepsilon}_1$ up to N$^3$LO, 
the deuteron quadrupole form factor  up to N$^2$LO and the isoscalar
amplitude in $np\rightarrow d\gamma$ up to N$^2$LO.


The lagrange density describing such interaction is~\cite{CRSa}
\begin{eqnarray}
  {\cal L}^{(sd)} & = &
-{\cal T}^{(sd)}_{ij,xy}\left(
  \CSDzero \left[P^i\right]^\dagger
  \left[ {\cal O}^{xy,j}_2 \right]
 + \CSDtwotwo
\left[ {\cal O}^{ll,i}_2  \right]^\dagger
  \left[{\cal O}^{ xy,j}_2 \right]
 +  \CSDtwoone
\left[P^i  \right]^\dagger
  \left[ {\cal O}^{mm, xy,j}_4 \right]
\right.
\nonumber\\
& & \left.
+  \CSDfour
\left[ {\cal O}^{mm,ll,i}_4  \right]^\dagger
  \left[ {\cal O}^{xy,j}_2 \right]
 +  \CSDfourt
\left[ {\cal O}^{ll,i}_2 \right]^\dagger
  \left[ {\cal O}^{mm,xy,j}_4 \right]
 + \CSDfourtt
\left[P^i  \right]^\dagger
  \left[{\cal O}^{aa,mm,xy,j}_6 \right]
\right)
\nonumber\\
& &
\ +\ ...\ \ 
\ +\ {\rm h.c.}
\ \ \ ,
\label{eq:sdlag}
\end{eqnarray}

where
\begin{eqnarray}
\left[ {\cal O} \right] & \equiv & \left( N^T {\cal O} N\right)
\ \ \ ,\ \ \ 
{\cal T}^{(sd)}_{ij,xy}
\ =\  
\left( \delta_{ix}\delta_{jy} - {1\over n-1} \delta_{ij}\delta_{xy}\right)
\nonumber\\
{\cal O}^{xy,j}_2 & = &
-{1\over 4}\left(
      \overleftarrow {\bf D}^x  \overleftarrow {\bf D}^y P^j
+ P^j \overrightarrow {\bf D}^x  \overrightarrow {\bf D}^y
- \overleftarrow {\bf D}^x P^j\overrightarrow {\bf D}^y
-\overleftarrow {\bf D}^y P^j\overrightarrow {\bf D}^x
\right)
\nonumber\\
{\cal O}^{wz,xy,j}_4 & = &
{1\over 16}\left(
      \overleftarrow {\bf D}^w  \overleftarrow {\bf D}^z {\cal O}^{xy,j}_2
+ {\cal O}^{xy,j}_2 \overrightarrow {\bf D}^w  \overrightarrow {\bf D}^z
- \overleftarrow {\bf D}^w {\cal O}^{xy,j}_2\overrightarrow {\bf D}^z
-\overleftarrow {\bf D}^z {\cal O}^{xy,j}_2\overrightarrow {\bf D}^w
\right)
\nonumber\\
{\cal O}^{ab,wz,xy,j}_6 & = &
-{1\over 64}\left(
      \overleftarrow {\bf D}^a  \overleftarrow {\bf D}^b {\cal O}^{wz,xy,j}_4
+ {\cal O}^{xy,j}_2 \overrightarrow {\bf D}^a  \overrightarrow {\bf D}^b
- \overleftarrow {\bf D}^a {\cal O}^{wz,xy,j}_4\overrightarrow {\bf D}^b
-\overleftarrow {\bf D}^b {\cal O}^{wz,xy,j}_4\overrightarrow {\bf D}^a
\right)
\ \ \ .
\label{eq:sdop}
\end{eqnarray}
We have not shown higher dimension operators, such as those corresponding to 
${\cal O}(p^8)$, but it is obvious how to include them and what the 
notation is.
The tree-level amplitude 
for an $\siii\rightarrow\diii$ transition
resulting from this Lagrange density is 
(in momentum space)
\begin{eqnarray}
{\cal A}^{\rm tree}_{sd} & = & 
-\left( \CSDzero + \left[\CSDtwotwo+\CSDtwoone\right] p^2
+\left[\CSDfour+\CSDfourt+\CSDfourtt\right] p^4\ +\ ...
\right)
\nonumber\\
& & \left[p^i p^j-{1\over n-1} p^2\delta^{ij}\right]
\ \ \ .
\label{eq:SDtree}
\end{eqnarray}
The coefficients that appear in 
eq.~(\ref{eq:sdlag})
themselves have an expansion in powers of $Q$,
e.g. $\CSDzero = \CSDzeromone + \CSDzerotwo+...$.

In order to calculate the $\siii-\diii$ mixing parameter 
$\overline{\varepsilon}_1$ up to N$^3$LO we only require one insertion
of ${\cal A}^{\rm tree}_{sd}$, dressed with the appropriate 
$\siii-\siii$ interactions.
The expression for $\overline{\varepsilon}_1$ is straightforward but long 
and so  we do not present it here.
We define the coefficients that arise in the momentum expansion of 
$\overline{\varepsilon}_1$, $E_1^{(2)}$ and $E_1^{(4)}$, by
\begin{eqnarray}
\overline{\varepsilon}_1 & = & E_1^{(2)} { p^3\over\sqrt{p^2+\gamma^2}}
\ +\ 
E_1^{(4)} { p^5\over\sqrt{p^2+\gamma^2}}
\ +\ ....
\label{eq:ep1exp}
\end{eqnarray}
The coefficients are fit to the Nijmegen partial wave
analysis\cite{nijmegen}, and are found to be
$E_1^{(2)}= 0.386 \ {\rm fm^2}$
and
$E_1^{(4)} \ =\  -2.800 \ {\rm fm^4}$.
As far as the power counting is concerned it is important to note that 
the coefficients 
$E_1^{(2)}$ and $E_1^{(4)}$ are set by physics at the high scale
and therefore are $Q^0$ or higher.
Thus, contributions of order $Q^{-l}$ must identically 
vanish\cite{FMSep}.
The superscript on the $E_1^{(n)\ \prime} s$ denotes the lowest order in $Q$ 
at which contributions may arise.
These conditions ensure non-trivial relations between the coefficients in
eq.~(\ref{eq:sdlag}) as the renormalization scale is reduced below the high
scale.
Each of the coefficients in eq.~(\ref{eq:sdlag}) can be written in terms of 
physical observables, such as $E_1^{(2)}$,  $E_1^{(4)}$, $\rho_d$, 
$\gamma$, and the renormalization scale $\mu$.

We are free to choose parameters other than the $E_1^{(n)\ \prime} s$
to expand in.  
A quantity that is
more directly related to the properties of the deuteron, 
is $\etasd$,
(written in terms of the $\siii-\diii$ mixing angle in the 
Blatt and Biedenharn parameterization of the S-matrix\cite{BBparam})
which is defined to be 
\begin{eqnarray}
\etasd & = & -\tan\left(\varepsilon_1\right)
\ \ \ ,\ \ \ 
\tan\left(2 \varepsilon_1\right)
\ = \ 
{ \tan\left(2 \overline{\varepsilon}_1\right)
\over
\sin\left(\overline{\delta}_0-\overline{\delta}_2\right)}
\ \ \ ,
\label{eq:asymDS}
\end{eqnarray}
evaluated at the deuteron pole, $|{\bf p}|=i\gamma$.
The difference between using $\{\etasd , E_1^{(4)} \}$ and 
$\{ E_1^{(2)} , E_1^{(4)} \}$ is higher order in the expansion.
In terms of the coefficients  $E_1^{(2)}$ and  $E_1^{(4)}$
it is easy to show that, up to N$^3$LO
\begin{eqnarray}
\etasd  & = & \gamma^2 \left[ 
E_1^{(2)} \left( 1 - {1\over 2}\gamma\rho_d - {1\over 8}\gamma^2\rho_d^2
-{1\over 16}\gamma^3\rho_d^3
\right)
- 
E_1^{(4)} \gamma^2\left( 1 -{1\over 2}\gamma\rho_d\right)
\right]
\ \ \ ,
\label{eq:sdasym}
\end{eqnarray}
which is, order by order,
$\etasd = 
0.0207 - 0.0042 + 0.0076 - 0.0017\ +\ ...
\ \ .$
Numerically, it is clear that the expansion is converging, 
but slowly due to the relatively large size of $E_1^{(4)}$ compared
to $E_1^{(2)}$.
This  slow convergence
will give rise to slow convergence in observables involving the deuteron
and therefore
it is convenient to invert this relation and use the very 
precise\cite{nijmegen} determination of $\etasd=0.02543\pm 0.00007$ as one of
the expansion parameters.  

\begin{figure}[t]
\centerline{{\epsfxsize=4.5in \epsfbox{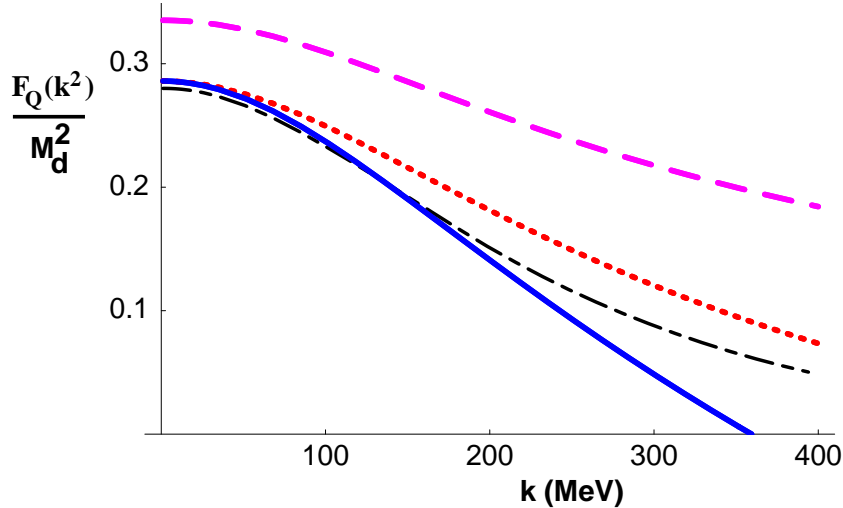}} }
\noindent
\caption{\it
The deuteron quadrupole form factor.
The dashed, dotted and solid  curves denote
the LO, NLO and N$^2$LO
deuteron quadrupole form factors computed in $\nopi$.
The dot-dashed curve corresponds to a 
calculation with the Bonn-B potential
in the formulation of  \protect\cite{AGA} .
}
\label{fig:Qplot}
\vskip .2in
\end{figure}

At LO in $\nopi$ the deuteron quadrupole moment is given entirely 
in terms\footnote{In \cite{CRSa} we wrote the quadrupole moment in terms of 
the $E_1^{(n)\ \prime} s$.  
This leads to slight numerical differences between the two
expressions (formally higher order differences).} of 
$\etasd$
\begin{eqnarray}
\mu_{\cal Q}^{(LO)} & = & {\etasd\over \sqrt{2}\gamma^2}
\ =\ 0.335\ {\rm fm^2}
\ \ \ .
\label{eq:quadmomLO}
\end{eqnarray}
At NLO the four-nucleon-one-photon operator\cite{CRSa}
with coefficient $\CQuad$ defined by the Lagrange density
\begin{eqnarray}
{\cal L} &=& { e\ \CQuad\over 2} \ \left( N^{T}P_{i}N\right) ^{\dagger }
\left( N^{T}P_{j}N\right)
\left[ \nabla^i{\bf E}^j\ +\  \nabla^j{\bf E}^i \ - \
  {2\over n-1}\delta^{ij} \ \nabla\cdot {\bf E}\right]
\nonumber\\
& = &  -e\ \CQuad \ \left( N^{T}P_{i}N\right) ^{\dagger }
\left( N^{T}P_{j}N\right)
\left( \nabla ^{i}\nabla ^{j}-\frac{1}{n-1}\nabla ^{2}\delta ^{ij}\right)
A^{0}
\ +\ ....
\ \ \ ,
\label{counterQ}
\end{eqnarray}
where ${\bf E}$ is the electric field operator,
contributes to the deuteron electric quadrupole moment as well as to the 
isoscalar E2 amplitude in $np\rightarrow d\gamma$.
The counterterm is determined by fitting the NLO amplitude to the observed
quadrupole moment, and it is convenient to define the 
quantity $\delta\mu_{\cal Q}$
\begin{eqnarray}
\delta\mu_{\cal Q} & = & \mu_{\cal Q}^{\rm expt.} - \mu_{\cal Q}^{(LO)}
\ =\ -0.0492\ {\rm fm^2}
\ \ \ ,
\label{eq:delQnum}
\end{eqnarray}
which is taken to scale as $Q^1$ in the power counting.
Solving for the quadrupole counterterm, one finds
\begin{eqnarray}
\CQuad & = & \pi {\sqrt{2}\ \etasd\ \rho_d - 2\  \delta\mu_{\cal Q}\  \gamma\over
2\gamma^2 (\mu-\gamma)^2}
\ \ \ .
\label{eq:quadsolve}
\end{eqnarray}
Naively, higher order quadrupole counterterms contribute to the
quadrupole form factor and quadrupole moment at N$^2$LO.
However, an RG analysis of such contribution shows that they first contribute
at N$^3$LO, and we can neglect them in our analysis.
Explicit calculation of the deuteron quadrupole form 
factor\cite{KSW2,CRSa} up to N$^2$LO
gives (neglecting relativistic corrections that are
suppressed by additional factors of $m_\pi^2/M_N^2$)
\begin{eqnarray}
{1\over M_d^2} F_{\cal Q} (|{\bf k}|) & = & 
\delta\mu_{\cal Q}
- 
{3\etasd \over 2\sqrt{2} \gamma |{\bf k}|^3}
\left[ \ 
4 |{\bf k}| \left( \gamma \left(1+\gamma\rho_d+\gamma^2\rho_d^2\right)
+{1\over 6} |{\bf k}|^2\rho_d \left(1+\gamma\rho_d\right)
-{1\over 6}\gamma r_N^2 |{\bf k}|^2\right)
\right.\nonumber\\
& & \left.
\qquad\qquad\qquad
-\left(3 |{\bf k}|^2+16\gamma^2\right)
\left( \left(1+\gamma\rho_d+\gamma^2\rho_d^2\right) 
-{1\over 6}r_N^2 |{\bf k}|^2\right)
\tan^{-1}\left({|{\bf k}|\over 4\gamma}\right)
\ \right]
\ \ \ ,
\label{eq:QFF}
\end{eqnarray}
and the deuteron quadrupole moment, $\mu_Q = F_{\cal Q}(0)/M_d^2$,
is reproduced straightforwardly.
$r_N=0.79\ {\rm fm}$ is the isoscalar nucleon charge radius that first enters
at N$^2$LO.
It is clear that up to N$^2$LO the quadrupole form factor has a well behaved
expansion in powers of $\gamma \rho_d$.
The overall normalization is largely determined by $\etasd$, with the
counterterm appearing at NLO required to reproduce the quadrupole
moment\cite{CRSa}.
It is interesting to note that even at N$^2$LO there is no contribution from
$E_1^{(4)}$, and the form factor is given entirely in terms of $\etasd$
$\gamma$, $\rho_d$ and  $\delta\mu_{\cal Q}$.
A plot of the quadrupole form factor at LO, NLO and N$^2$LO can be found in 
fig.~(\ref{fig:Qplot}).
%
%


Using the above analysis we are in a position to make a prediction 
for the isoscalar amplitude in $np\rightarrow d\gamma$ up to N$^2$LO.
The LO, NLO and N$^2$LO contributions to  $\tilde{X}_{E2_{S}}$ are 
\begin{eqnarray}
\tilde{X}_{E2_{S}}^{(LO)}  & = &  -{1\over 10}\ \etasd
\ \ =\ \ -2.54\times 10^{-3}
\nonumber\\
\tilde{X}_{E2_{S}}^{(NLO)} & = & -{3 \gamma\rho_d\over 80} \ \etasd\ +\ 
{\gamma^2\over 4\sqrt{2}}\ \delta\mu_{\cal Q}
\ \ =\ \ -0.86\times 10^{-3}
\nonumber\\
\tilde{X}_{E2_{S}}^{(N^2LO)} & = & -{\gamma^2\rho_d^2\over 64} \ \etasd
\ +\ {3\gamma^4 \over 40}\ E_1^{(4)}
\ \ =\ \ -0.67\times 10^{-3}
\ \ \ .
\label{eq:E2LO}
\end{eqnarray}
It is clear that the perturbative expansion is converging, however, the 
ratio of the third to second term in the expansion is not 
particularly small.  This suggests that a N$^3$LO
calculation is required before one has confidence in the value of this
amplitude.  A conservative estimate of the uncertainty in the N$^2$LO
calculation is the size of the N$^2$LO contribution itself.


Numerically, the $\nopi$ calculations of the subleading amplitudes for
near threshold
$np\rightarrow d\gamma$ capture are\footnote{Our definition of the $E2_S$ amplitude is of opposite sign 
to that used in \cite{PMRsupp}, 
and hence ${\cal R}_{\rm E2}~=~-~\tilde{X}_{E2_{S}}/
\tilde{X}_{M1_{V}}$}
\begin{equation}
\frac{\tilde{X}_{M1_{S}}}{\tilde{X}_{M1_{V}}}=-5.0\times 10^{-4}
\qquad\qquad , \qquad\qquad
\frac{\tilde{X}_{E2_{S}}}{\tilde{X}_{M1_{V}}}\ =\  
-2.5\times 10^{-4}
\label{eq:M1num}
\end{equation}
with uncertainties that we naively estimate to be of order  $\sim 60\%$
and $\sim 15\%$ respectively, due to the omission of higher order terms.
For an incident  neutron speed of
$\left| {\bf v}\right|~{\rm m/s}$ in the
proton rest frame, we find
\begin{equation}
\frac{\tilde{X}_{E1_{V}}}{\tilde{X}_{M1_{V}}}
\ =\ -1.2\times 10^{-4}\ \left({ |{\bf v}|\over 2200}\right)
\quad ,
\label{eq:E1num}
\end{equation}
with an uncertainty that we estimate to be of order  $\sim 3\%$\cite{CSbb}.
Even for neutrons with  $\left| {\bf v}\right|=2200~{\rm m/s}$ the 
$E1_V$ capture
cross section is comparable to the suppressed amplitudes for $M1_S$ 
and $E2_S$ capture.

Using these amplitudes to compute the photon polarizations  $P_\gamma$,
we find
\begin{eqnarray}
P_\gamma (M1)& =& -7.1\times 10^{-4}
\ \ \ ,\ \ \ 
P_\gamma (E2) \ =\  -3.5\times 10^{-4}
\ \ \ ,
\label{eq:polnums}
\end{eqnarray}
giving a total of
$P_\gamma =-1.06\times 10^{-3}$ in the forward direction, 
approximately $2/3$ of the
experimentally determined value of\cite{Bazh}
$P_\gamma^{\rm expt} =-(1.5\pm 0.3) \times 10^{-3}$.
Given the large uncertainty in the calculation of the $M1_S$ amplitude,
and the uncertainty of the measurement, the two are consistent at the order
to which we have calculated.
Our value of $P_\gamma (M1)= -7.1\times 10^{-4}$ is in complete agreement
with the results of Burichenko and Kriplovich\cite{BKa}
of $P_\gamma (M1)= -7.0\times 10^{-4}$ from a Reid soft-core calculation,
but is somewhat less than their zero-range calculation of
$P_\gamma (M1)= -9.2\times 10^{-4}$.
However, given the large uncertainty in our $M1_S$ amplitude, both values are
consistent.
Our value of $P_\gamma (E2)  =  -3.5\times 10^{-4}$ agrees well\footnote{Given 
that our NLO calculation reproduces the numerical value
of both the $M1_S$ and $E2_S$ amplitudes computed 
by Park, Kubodera, Min and Rho\cite{PMRsupp},
the ``Rho-Challenge'' has been met.  
These observables do not distinguish between the two approaches 
at the order to which we are working.
}
with the 
recent calculation of Park, Kubodera, Min and Rho\cite{PMRsupp}, and lies
somewhere between the zero-range approximation calculation of
$P_\gamma (E2)  =  -2.4\times 10^{-4}$
(which we reproduce at LO in $\nopi$)
and Reid soft-core  calculation of 
$P_\gamma (E2)  =  -3.7\times 10^{-4}$ 
by Burichenko and Kriplovich\cite{BKa}.

The power of effective field theory is that there are well-defined 
expansion parameters, even when loop graphs appear.
It is therefore natural to understand the power counting of the 
spin-dependent asymmetries that we have considered.
The $M1_S$ amplitude starts at order $Q^0$, but receives its first non-zero 
contribution at order $Q^1$.  We have only computed the order $Q^1$ contribution.
In contrast, the $E2_S$ amplitude starts at order $Q^2$
and we have computed the order $Q^2$, $Q^3$ and $Q^4$ contributions.
Therefore, the observable $P_\gamma$ has been computed only to order $Q^1$, 
despite our calculation of part of the order $Q^2$, $Q^3$ and $Q^4$
contributions from the $E2_S$ amplitude.  
This is apparent in the size of the uncertainty
arising from higher order terms in the $M1_S$ amplitude, 
that we have discussed
extensively.
A similar statement can be made about the angular asymmetry, 
in particular $S^{(2)}$, which starts at order $Q^2$ with 
the interference between $M1_S$ and $E2_S$ starting at $Q^3$.
Experimentally, measurement of both asymmetries will allow for an 
extraction of both $M1_S$ and $E2_S$ 
(when $E1_V$ is negligible and noting that
the amplitudes are real at threshold), 
as is clear from eq.~(\ref{eq:sdef}) and eq.~(\ref{eq:pdef}).

In conclusion, 
we have used the effective field theory without  pions that describes
the nucleon-nucleon interaction to find analytic expressions for
the isoscalar $M1$ and  isoscalar $E2$
contributions to the $np\rightarrow d\gamma$ capture process near
zero incident nucleon momentum.
The $E2_S$ amplitude
is determined at the $15\%$ level, and we find a value consistent
with previous calculations.
Due to the vanishing contribution of the one-body operator up to
NLO, the uncertainty in the $M1_S$ amplitude is
estimated to be at the $60\%$ level.
This relatively large uncertainty at NLO is consistent with the range of
amplitudes determined with other approaches.
A N$^2$LO calculation may be able to reduce this uncertainty.
However, additional counterterms that may arise at N$^2$LO 
must be determined elsewhere, otherwise more precise
predictions for these subleading amplitudes will not be possible.

\vskip 0.5in

We would like to thank Mannque Rho for challenging us to make estimates of
these suppressed amplitudes.
We would also like to thank Jim Friar for a question he asked us two years
ago about the prediction for $\etasd$ in effective field theory.
This work is supported in part by the U.S. Dept. of Energy under Grants No.
DE-FG03-97ER4014.

\end{document}